\documentclass{neuthist18}
\def\Journal#1#2#3#4{{#1} {\bf #2}, #3 (#4)}
\def\NCA{\em Nuovo Cimento}

\def\PLB{{\em Phys. Lett.}  B}
\def\PRL{\em Phys. Rev. Lett.}
\def\PRD{{\em Phys. Rev.} D}

\def\be{\begin{equation}}
\def\ee{\end{equation}}
\def\bea{\begin{eqnarray}}
\def\eea{\end{eqnarray}}

\newcommand{\Photo}{\includegraphics[height=35mm]{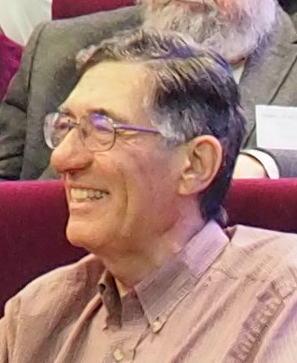}}

\begin{document}
\vspace*{4cm}
\title{The Discovery of the Atmospheric Neutrino Anomaly}
\author{ John M. LoSecco }
\address{Physics Department, University of Notre Dame du Lac, Notre Dame,
  Indiana 46556-5670, USA}
\maketitle\abstracts{
  The discovery of the missing atmospheric muon neutrinos~\cite{Pers},
  known as the atmospheric neutrino anomaly, is briefly described.
  Learned and Lipari gave a general review of atmospheric neutrinos at the conference,
  including the discovery of the anomaly by IMB-1 and subsequent work.  
Questions answered in this brief note include: the cautious wording, the statistical
significance, the 1992 erroneous exclusion plot, the Kamiokande confirmation and
Super-Kamiokande's failure to cite the original 1986 IMB-1 discovery.}
\section{Preface}
The detailed story of the discovery of the atmospheric neutrino anomaly has
been told before~\cite{Pers}.  Due to length limitations in these proceedings,
this article will concentrate on responding to questions that have been raised
about the discovery and subsequent work to understand the physical origin
of the anomaly.  In their talks at the conference Paolo Lipari and John Learned reviewed the
discovery as told in reference~\cite{Pers}.
\section{Synopsis of Reference 1}
The goal of the  IMB-1 (Irvine-Michigan-Brookhaven) experiment was to discover proton decay.  It was expected that the dominant background would come from atmospheric neutrino interactions.
Estimated neutrino fluxes and cross sections were used to simulate this expected
background.  Reconstruction of events in the experiment was based on the flight time
of Cerenkov photons.  A modest modification to the timing circuits let us record
activity for 10 $\mu$sec after a trigger.  This gave a delayed signal in events
containing a muon.  The efficiency of the method was determined by observing
stopping cosmic ray muons from the surface, and agreed very well with expectations.

From the very beginning the IMB-1 experiment measured fewer muon decay events than
expected from atmospheric neutrinos.  26$\pm$2\% of the events were observed to
have a muon decay while the expected value was 34$\pm$1\%.
Numerous checks were performed to determine the detector response to muons was
well understood.  The expected value was studied by varying the production model
using explicit $\nu_{\mu}$ interactions on $CF_{3}Br$, neon and deuterium as well
as the Rein and Seghal model of neutrino interactions~\cite{ReinSehgal}.

The muon
deficiency was published in several PhD thesis~\cite{IMB1983}, a couple of conference
proceedings~\cite{LL} and a Physical Review Letter~\cite{IMB2}.  The February 1986
Lake Louise proceedings~\cite{LL} explicitly noted that IMB had measured $\nu_{e}/\nu_{\mu}$=1.3 while at that time Nusex and Kamiokande were reporting
$\nu_{e}/\nu_{\mu}$=0.28$\pm$0.11 and $\nu_{e}/\nu_{\mu}$=0.36$\pm$0.08 respectively.
The IMB evidence was strong, $3.5 \sigma$,
but confirmation was needed.  Shortly after the 1986 PRL article had been submitted, I asked Kamiokande to confirm the anomaly, pointing out a muon decay deficiency in their own data.
After a substantial delay, while they redesigned their particle classification
algorithm, confirmation was provided~\cite{Conf}.

\section{Summary of the Poster}
\begin{figure}[h]
\centerline{\includegraphics[width=0.95\linewidth]{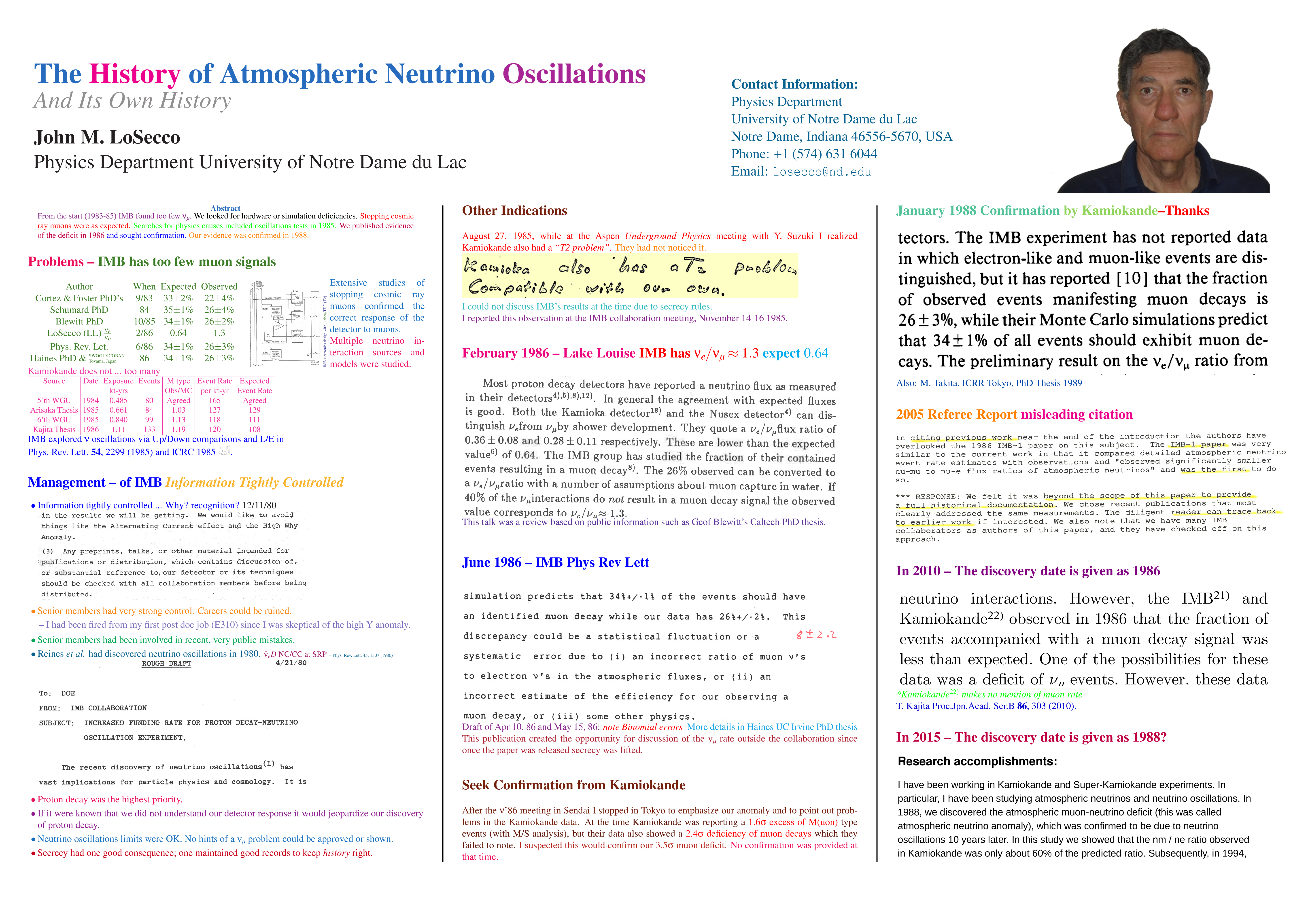}}
\caption[]{Summary of the published history article~\protect\cite{Pers}.The full size copy of the poster is available at the address http://neutrinohistory2018.in2p3.fr/programme.html, for easier reading.}
\label{poster}
\end{figure}

	\noindent
	In general, the poster (see Fig.~\ref{poster}) summarized the published history
	article~\cite{Pers} with a few additional details.  The additions included a
	discussion of the IMB management and biases (lower left), my realization in August 1985 that
	Kamiokande also had a muon decay deficiency (upper center) and a brief survey of how some of the
	history has been overlooked by others to promote followup work (lower right).
These additions had been left out of reference~\cite{Pers} for two reasons. Reference~\cite{Pers} was celebration of a colleague's career and such negative material would not have been appropriate. Also, while well documented, the information in these additions comes from private archives of the IMB experiment which has had limited public access.
	
	The IMB collaboration had a secrecy rule to limit rumors of the expected
	discovery of proton decay.  The rules were enforced by a senior management team that had a record of prior mistakes. 
 
	At the ICRC in 1985 Nusex showed evidence of an excess $\nu_{\mu}$ rate as determined by comparing showering and non-showering tracks in an iron calorimeter.  This was confusing since IMB had solid evidence that we were seeing too few muon decays.  A few months later I realized that Kamiokande
	had information that spanned both possibilities.   That $\mu$ $e$ based pattern
	identification supporting the Nusex result but I noticed that their muon decay
	was compatible with the IMB anomaly.  I could not discuss this at the time
	due to IMB secrecy rules.
	
The blue text at the bottom of the upper left section of the poster shows the atmospheric neutrino E/L plot from the 1985 ICRC where the discrepancy, a dip in the bin centered at E/L\,=\,5.8$\times$10$^{-3}$ MeV/meter, was clearly noted in the original.

\section{Cautionary Wording}
In his talk at the conference Paolo Lipari was critical of the language used in the
first IMB journal article~\cite{IMB2} to explicitly mention the atmospheric muon deficit.

The Physical Review Letter~\cite{IMB2} did not give a strong interpretation to the
missing muons.  Most of the collaboration was quite cautious to claim neutrino
oscillations for several reasons.  The text~\cite{IMB2} read ``This discrepancy could be a
statistical fluctuation or a systematic error due to (i) an incorrect ratio of muon
$\nu$'s to electron $\nu$'s in the atmospheric fluxes, or (ii) an incorrect estimate of the efficiency for our observing a muon decay, or (iii) some other
as-yet-unaccounted-for physics.''  Which makes explicit what any cautious reader should be thinking.

The atmospheric neutrino flux calculations were not our own.
There was no guarantee that all of the data events were caused by neutrinos
of atmospheric origin but that was the model to which we compared the sample.  An efficiency check with cosmic rays worked because the $\mu^{-}$ to $\mu^{+}$ ratio of that sample was known.  So we knew it was not a detector problem.

At the time of the IMB publication two other experiments sensitive to atmospheric neutrinos,  Nusex and Kamiokande, using different methods, were reporting an excess of muon type events in their data samples.  This is why IMB's observation was a discovery.  We were the first to report the correct value.

I had provided an interpretation of the anomaly in an earlier conference paper~\cite{LL} as a $\nu_{e}$ to
$\nu_{\mu}$ ratio of 1.3, with a muon detection efficiency of 60\%.

Earlier attempts to fit the neutrino oscillations hypothesis~\cite{IMB1,ICRC} to the
data only placed limits since the oscillation parameters were not in a range 
to which the experiment was sensitive.  Those attempts were motivated by the,
at the time unpublished, anomaly.

Many senior authors were scared by the anomaly and there was a strong bias against neutrino oscillations due to prior mistakes on other projects.  To get the
correct result published by the collaboration required patience, great attention to detail, redundancy and tenacity.
\section{Significance}
The statistical significance of the evidence published by IMB in 1986 was 3.5 $\sigma$.  The published
observed value of the muon decay fraction was 26$\pm$3\%. Drafts of the paper distributed to the collaboration for approval had the correct 26$\pm$2\%.
The $\pm$2 is based on binomial statistics.  An event had a muon decay or it did not.  I do not know how the $\pm$3 got in the paper.
\section{Later Papers -- The 1992 IMB Exclusion Plot}
In his talk at the conference Maury Goodman mentioned an IMB paper~\cite{NoOsc} from 1992
that ruled out regions of neutrino oscillations parameter space that are now
believed to be the correct physical ones.  This was my response at the meeting.

After the discovery (1986) and confirmation (1988) work on atmospheric neutrinos intensified.
Many people and groups joined in.  IMB-3 had 4 times the light collection as IMB-1
and several pattern based muon-electron discrimination algorithms were developed
(and checked against the observed muon decay rates).
To prove neutrino oscillations one needed to show clear evidence of an L/E dependence.  While atmospheric neutrinos have flight paths from a few km to 12,700 km due to the modest solid angle near the horizon the path length distribution
is dominated by two distance scales, dozens of km for the downward component
and about 10,000 km for the upward.  The neutrino energy seems to be predominately
below 1 GeV.  One must be creative to extend the range of energies one can observe.

To get more events at higher energies required a much larger detector.  Resources
for a larger detector were not available but if one only needed the larger detector to observe higher energies one can utilize neutrino interactions in the rock below these underground detectors.  The paper in question~\cite{NoOsc} used the fraction of stopping upward going muons, relative to upward going muons that exit the detector, to constrain neutrino oscillations.  Upward going muons are caused by
neutrino interactions in the rock surrounding the detector.

While a very nice idea it must be cautiously executed.  There are no reliable
energy estimates for entering tracks so one is integrating over a broad range of
$\nu_{\mu}$ energies assuming the theoretical spectrum has been calculated correctly.
The stopping fraction should be insensitive to the flux normalization since it is the ratio of two parts of the same spectrum.  But in this case~\cite{NoOsc} there
was no single flux estimate that could span the range of neutrino energies needed
so two different flux estimates were used:  Volkova~\cite{Volkova} for the high energy part and Lee and Koh~\cite{LKFlux} for the low energy part.  The Lee and Koh
flux~\cite{LKFlux} was later shown to be wrong due to a programming bug~\cite{LKBad},
but this was not realized until years later in a general review of all atmospheric
flux estimates.  It underestimated the low energy flux, which made its prediction
look more like the correct flux with neutrino oscillations.

\section{Independent Discovery?}
As mentioned above, in June 1986 I asked Kamiokande to confirm our evidence of a muon
deficit.  At the time they were reporting a 1.6$\sigma$ muon excess.  At the 2018 conference, I asked Takaaki Kajita to confirm the time line.  In his response he
indicated that Kamiokande had relied on scanning to classify events as muon or electron.  It wasn't until Fall of 1986 that an automated method was developed.
This confirms what is indicated in Takita's 1989 PhD thesis~\cite{Takita}. But the thesis doesn't give specific dates.
Kajita indicated there had been no formal particle identification in Kamiokande
before I mentioned the muon deficit.  The Kamiokande work was not independent. It was
a confirmation of the IMB result.

\section{Apology}
At the meeting, I asked Kajita why Kamiokande had cited the earlier 1986 work
from IMB but that Super-Kamiokande never cited the earlier IMB paper.  His immediate reply was ``I'm sorry''~\cite{NuHK}.  He went on to explain that Super-Kamiokande cited later IMB-3 papers~\cite{IMB3}
that used pattern based particle identification methods.  All IMB-3 contained atmospheric neutrino comparisons are inaccurate since they were modeled with the Lee and
Koh neutrino spectrum~\cite{LKFlux} which is flawed by a programming bug which underestimates the muon neutrino flux.

\section{Video of Conference Discussions}

The talks and discussions at the conference were recorded and are available at the address:
\\ http://neutrinohistory2018.in2p3.fr/programme.html. Many of the new material in this article is documented in recordings of discussions.

\section*{Acknowledgments}
I would like to thank the conference organizers for inviting me to participate.
In particular Daniel Vignaud and Michel Cribier were quite adaptable to my requests.  I thank the other participants for making it a lively meeting and for their candor.
I thank the physics department at Notre Dame for their partial support.

\end{document}